\documentclass[aps,pre,10pt,twocolumn]{revtex4-1}
\usepackage{graphicx}
\usepackage{amsmath}
\usepackage{mathrsfs}
\usepackage{graphicx}
\usepackage{epstopdf}
\usepackage{amsmath}
\usepackage{amssymb}
\usepackage[english]{babel}
\newcommand{\be}{\begin{equation}}
\newcommand{\bea}{\begin{eqnarray}}
\newcommand{\bc}{\begin{center}}            
\newcommand{\ee}{\end{equation}}
\newcommand{\eea}{\end{eqnarray}}
\newcommand{\ec}{\end{center}}

\newcommand{\baa}{\begin{eqnarray*}}
\newcommand{\eaa}{\end{eqnarray*}}
\begin{document}
\title{Irreversible thermodynamics of thermoelectric devices: \\From local framework to global description}
 \author{Jasleen Kaur and Ramandeep S. Johal} 
 \email[e-mail: ]{rsjohal@iisermohali.ac.in}
 \affiliation{Indian Institute of Science Education and 
 Research Mohali, Department of Physical Sciences, 
 Sector 81, S.A.S. Nagar, Manauli PO 140306, Punjab, India}
\begin{abstract}
Thermoelectricity is traditionally explained via Onsager's irreversible, flux-force 
framework. The coupled flows of heat and electric charge are modelled as steady-state
flows, driven by the thermodynamic forces defined in terms
of the gradients of local, intensive parameters like temperature and electrochemical
potential. A thermoelectric generator is a device with a finite extension, and 
its performance is measured in terms of total power output and total entropy
generation. These global quantities are naturally expressed in terms of
discrete or global forces derived from their local counterparts. 
We analyze the thermodynamics of thermoelectricity in
terms of global flux-force relations. These relations clearly
show the additional quadratic dependence of the driver flux on global forces,
corresponding to the process of Joule heating.
We discuss the global kinetic coefficients defined by these flux-force relations
and prove that the equality
of the global cross-coefficients is derived from a similar property of the 
local coefficients. Finally, we clarify the differences between
the global framework for thermoelectric energy conversion and the recently
proposed minimally nonlinear irreversible thermodynamic model. 
\end{abstract} 
\maketitle
\section{Introduction}
In many transport phenomena, interesting effects arise
due to coupling between different
thermodynamic forces present simultaneously in the system. 
Thermoelectricity provides a physically transparent picture of this coupling 
between flows of heat and electric charge, which 
gives rise to the well-known Seebeck, Peltier, and Thomson effects 
\cite{Onsager1931, Callen1948, Domenicali1954, Callenbook1985, DavidJou2008}.
These interference phenomena
can be successfully described within the linear flux-force formalism
of Onsager \cite{Onsager1931}. The aforesaid linear relationship 
between fluxes and forces derives 
from the usually small magnitudes of the thermodynamic forces driving the system. 
The latter are expressed as gradients of the locally defined
intensive parameters and this framework invokes the 
local-equilibrium hypothesis \cite{Callen1948}.

Although, the thermodynamic explanation of the above phenomena
treats them as steady-state processes at the local level,
actual devices have macroscopic extensions and so their 
performance needs to be analyzed by scaling up the local description.
This leads to the study of global quantities like total power output
and total entropy generation by the device.
Due to the increasing worldwide demand for efficient and environment-friendly 
energy convertors, many works have pursued optimization 
of the material properties as well as power output/cooling power
of thermoelectric devices  \cite{Gordon1991, DiSalvo1999,  RIFFAT2003, Shakouri2011, 
Benenti2011,snyder2011, Goupil2011, Apertet2012B, Apertet2013A, Ouerdane2013, 
Wohlman2014,BENENTI2017,Jasleen2019, Ding2019}. It is noteworthy 
that a local, linear-irreversible model gives rise to 
nonlinear (quadratic) dissipation terms in the flux equations at the global level \cite{CWelton}.
In thermoelectricity, Joule heating plays the role of this 
dissipation term. Thus, despite the apparent presence of nonlinearities 
in thermal flux equations, the locally linear character of the underlying framework
for thermoelectricity has been emphasized  \cite{CWelton, Apertet2013B}.

With the advent of finite-time thermodynamics, the flux-force framework
for irreversible phenomena has been extended to macroscopic
heat devices too \cite{Caplan1965, VandenBroeck2005, Iyyappan2020}, where the 
fluxes still are linear functions of the {\it global} or discrete forces. 
Further, the so-called minimally nonlinear irreversible thermodynamic 
(MNLIT) model \cite{Izumida2012, Izumida2015} proposes generalized heat 
flux equations by incorporating a nonlinear, phenomenological
term at the global level, akin to the dissipation term in thermoelectric devices. 
This model may be adapted for both autonomous as well as cyclic heat devices 
operating in finite time \cite{Esposito2010, Izumida2013}. 
 
In this paper, our focus is on the thermoelectric
flux-force relations at the global level. 
As with the standard Onsager-Callen 
framework, consideration of the total rate of entropy generation
as a bilinear form leads to the identification of global thermodynamic forces.
We show how the scale-up from a local description naturally leads us
to express the {\it driven} flux as a linear function of these global or discrete forces.  
In consequence, we will see that the {\it driver} flux additionally follows a nonlinear
dependence on these forces, equivalent to the presence
of a nonlinear dissipation term discussed above.
We also define global kinetic coefficients from these flux relations. 
A hallmark of the underlying local framework is the equality of kinetic 
cross-coefficients based on principle of microscopic time reversibility \cite{Onsager1931b}.
We observe a similar property of the global kinetic coefficients,
which can be traced to its local counterpart. 

The plan of the paper is as follows. In Section II, 
Onsager-Callen framework for a thermoelectric generator
is discussed. In Section III, we identify the discrete or global forms
of forces and express the fluxes in terms of these forces. 
We also obtain expressions for the global kinetic coefficients and 
prove Onsager reciprocity at the global level. 
In Section IV, our approach is compared with the MNLIT model.  
Finally, we conclude in Section V.
\section{Thermoelectric generator}
Consider the thermoelectric material 
 to be a one-dimensional, homogeneous
 substance of length $l$ and cross-sectional area $A$, 
 with given values of electrical resistivity $\rho$, 
 thermal conductivity $\kappa$, 
and Seebeck coefficient $\alpha$.
This is the 
so-called Constant Properties model \cite{Ioffe1957}.
The two arms of the thermoelectric module are in series
electrically, while they are parallel to each other thermally (see Fig. 1).
\begin{figure}
	\includegraphics[width=6cm]{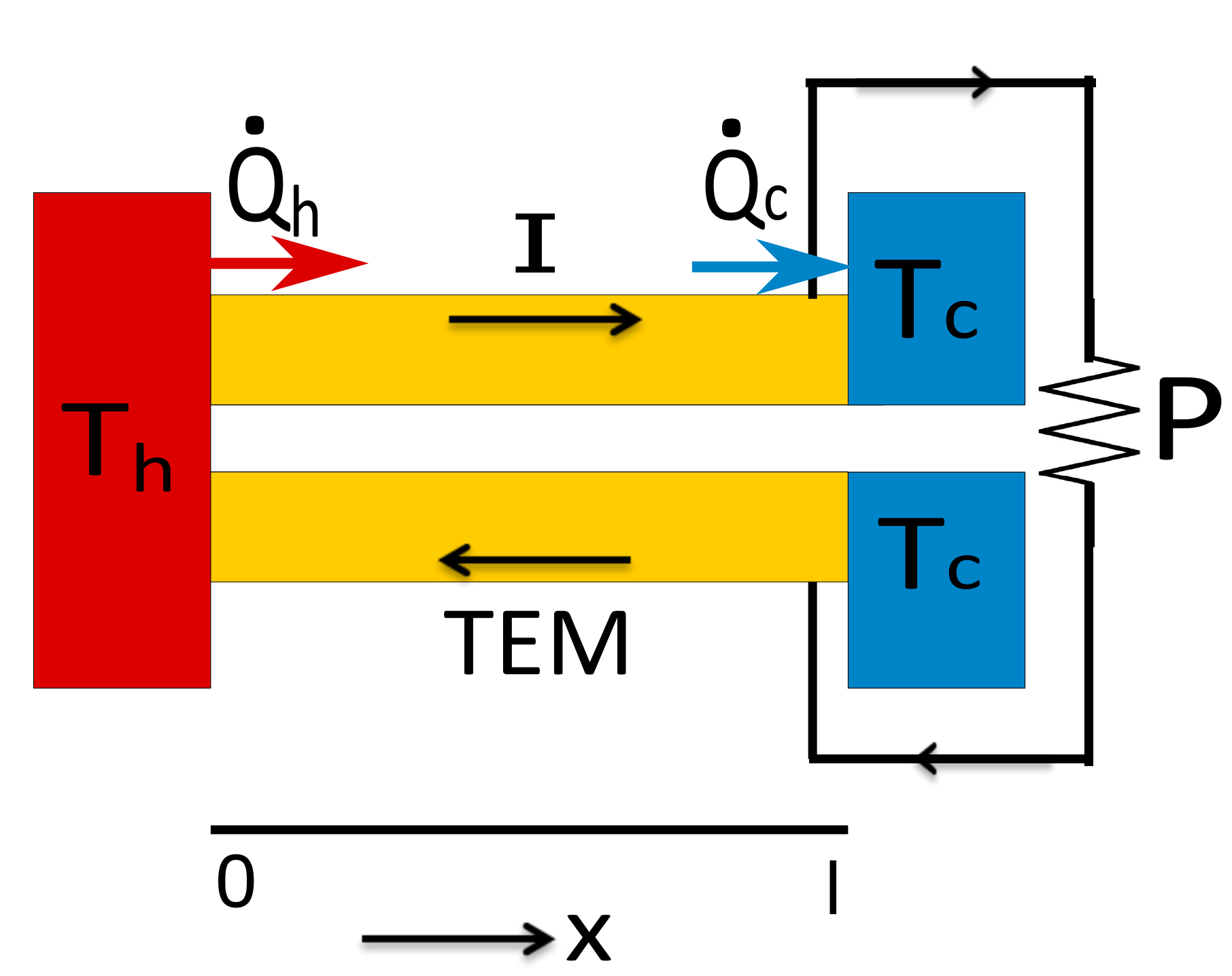} 
\caption{Schematic of thermoelectric heat engine. Two legs (yellow online)
of a constant properties thermoelectric material (TEM) in simultaneous contact with a hot
 ($x=0$) and a cold ($x=l$) reservoir.
Electric power, $P = \dot{Q}_h - \dot{Q}_c$,  
is drawn against an external load. }
\label{fig1}
\end{figure}
To describe thermoelectricity within 
linear-irreversible thermodynamic framework, 
the fluxes may be chosen as the 
particle current density $\vec{J}_N$ and the thermal 
current density $\vec{J}_Q^{}$, locally driven by 
the respective affinities $-\vec{\nabla} \mu /T$ and $\vec{\nabla} (1/T) = -\vec{\nabla} T/T^2$, where
$\mu$ and $T$ are the local electrochemical potential
and the temperature, respectively.
Assuming small magnitudes for such affinities, Onsager expressed  
the local fluxes as linear combination of these affinities:
\be           
\left(
\begin{array}{c}
 \vec{J}_N \\
\vec {J}_Q
\end{array}
\right)
=
\left(
\begin{array}{cc}
 L_{11} &   L_{12} \\ 
 L_{21} &   L_{22}
 \end{array}
\right)
\left(
\begin{array}{c}
 -{\vec{\nabla} \mu}/{T} \\
 -\vec{\nabla} T/{T^2}
\end{array}
\right), 
\label{ons1}
\ee
where the Onsager coefficients, $L_{ij}^{}$, obey certain conditions 
\cite{Onsager1931} in order to satisfy the second law of thermodynamics.
By defining the electric current density $\vec{J} = e \vec{J}_N$, 
and noting that $\vec{\nabla} \mu = e \vec{\nabla} V$,
where $e$ is the charge on electron and $V$ is 
the electrostatic potential, we can write 
(\ref{ons1}) as
\be           
\left(
\begin{array}{c}
\vec{ J }\\
\vec{J}_Q
\end{array}
\right)
=
\left(
\begin{array}{cc}
 e^2 L_{11} &   e L_{12} \\ 
  e L_{21}  &   L_{22} 
 \end{array}
\right)
\left(
\begin{array}{c}
 -{\vec{\nabla} V}/{T} \\
 -\vec{\nabla} T/{T^2}
\end{array}
\right).
\label{ons2}
\ee
It is known that the knowledge of Onsager coefficients $L_{i j}$
is equivalent to the knowledge of the quantities
$\rho$, $\kappa$ and $\alpha$, whereby
we have the following expressions \cite{Callen1948}: 
\bea
L_{11} &=& \frac{T}{\rho e^2}, \nonumber \\ 
L_{12} &=& \frac{\alpha T^2}{\rho e} = L_{21}, \label{lij} \\
L_{22} &=& T^2 \left(\frac{\alpha^2 T}{\rho} + \kappa \right).\nonumber 
\eea
Note that the cross-coefficients are equal 
by virtue of the principle of microscopic reversibility \cite{Onsager1931}.
Secondly, even if the quantities $\rho$, $\kappa$, and $\alpha$
are assumed to be constant, the Onsager coefficients
depend on local temperature $T$.
\section{From local to global affinities}
Now, the Onsager-Callen framework was formulated 
for the steady-state description
of irreversible phenomena in terms of forces and fluxes
at the {\it local} level. In this section, we analyze how 
the flux-force relations are extended to the macroscopic
level i.e. applied to the full length of the thermoelectric material.

The local rate of entropy production per unit volume, $\dot{s}$, 
is defined as the divergence of the entropy flux $\vec{J}_s = \vec{J}_Q/T$,
so that $\dot{s} = \vec{\nabla}. (\vec{J}_Q/T)$, which can be 
written as the sum of products of 
fluxes and conjugate forces, as follows.
\be 
\dot{s} = \vec{J} .\left( \! - \frac{\vec{\nabla} V}{T} \!  \right) + \vec{J}_{Q}^{}(x).
\vec{\nabla} \left( \!  \frac{1}{T} \! \right),
\label{sdot}
\ee
where we have used \cite{Callen1948}
\be
\vec{\nabla}.\vec{J}_Q = -\vec{J}.\vec{\nabla} V.
\label{jqjv}
\ee
Then, the rate of total entropy production per unit area (p.u.a) along
the whole length of the material (1-d case) is 
$\dot{\cal S} = \int_{0}^{l} \dot{s}\; dx$, and given by
\be
\dot{\cal S} = \frac{J_{Q}^{}(l)}{T_c} - \frac{J_{Q}^{}(0)}{T_h},
\label{sdot2}
\ee
where $x=0 (l)$ corresponds to the hot (cold) end
of the material.\\

Now, the global power output p.u.a is given by
\begin{eqnarray}
{\cal P} &= & -\int_{0}^{l} \vec{\nabla}. \vec{J_Q}\; dx, 
\label{calp1} \\
 &=& J_Q(0)-J_Q(l). \label{calp2}
\label{P1}
\end{eqnarray}
Further, the use of Eq. (\ref{jqjv}) in (\ref{calp1}),  
and the constant magnitude $\vert \vec{J}\vert \equiv J $ implies that 
\be
{\cal P}= J(V_c-V_h) \equiv J\Delta V.
\label{P2}
\ee
From Eqs. (\ref{P1}) and (\ref{P2}), we can write Eq. (\ref{sdot2}) as
\be
\dot{\cal S} = J \left( \! - \frac{\Delta V}{T_c} \!  \right)
  + J_{Q}^{}(0) \left(\!  \frac{1}{T_c} - \frac{1}{T_h} \! \right).
  \label{sdotmac}
\ee
Thus, the global rate of entropy production p.u.a 
is expressed in a bilinear form, 
$\dot{\cal S} \equiv {\cal J}_1 X_1 + {\cal J}_2 X_2$, where   
 the  corresponding flux-force pairs are identified as follows.
\bea
{\cal J}_1 &=& J, \qquad \quad X_1 = - \frac{\Delta V}{T_c}, \\
{\cal J}_2 &=& J_{Q}^{}(0), \quad X_2 = \frac{1}{T_c} - \frac{1}{T_h}.
\eea
Comparing Eqs. (\ref{sdot}) and (\ref{sdotmac}), we observe 
how the pair of local affinities take up a discrete form,  
relevant for the global description.
It is interesting
to analyze this bilinear form further from 
irreversible thermodynamic point of view. 
For future comparison, we consider 
the rate of total entropy production, $\dot{S} = 
\dot{\cal S} A  \equiv {J}_1 X_1 + {J}_2 X_2$, where 
\bea
{J}_1 &=& I, \qquad X_1 = - \frac{\Delta V}{T_c},
\label{j1mac}\\
{J}_2 &=& \dot{Q}_h, \quad X_2 = \frac{1}{T_c} - \frac{1}{T_h},
\label{j2mac}
\eea
with $I = J A$ and $\dot{Q}_h^{} = J_Q(0) A$.

\subsection{Flux-force relations at the global level}
Corresponding to linear flux-force relations
at the local level, we now enquire into the 
relations between fluxes and forces  
at the macroscopic level.
Note that the local
flux-force relations are {\it postulated} to be  
linear within Onsager's approach. Here, we wish to see   
how the fluxes, $J_i$, in Eqs. (\ref{j1mac}) and (\ref{j2mac}), 
are expressed in terms of the global or discrete affinities, $X_i$.

Firstly, from Eqs. (\ref{ons2})
and (\ref{lij}), we can write 
\be
\vec{J} = -\frac{1}{\rho}\vec{\nabla} V - \frac{\alpha}{\rho} \vec{\nabla} T.
\label{jet}
\ee
For the case of a 1-d thermoelectric element, 
we have
\be
J = -\frac{1}{\rho} \frac{\partial V}{\partial x} 
- \frac{\alpha}{\rho} \frac{\partial T}{\partial x}.
\label{jet1}
\ee
Now, since the current density $J$ as well as 
$\rho$ and $\alpha$ are constant 
along the length of the material,
integrating the above equation over $x \in [0,l]$,
we obtain:
\be
J.l = -\frac{\Delta V}{\rho} + \frac{\alpha}{\rho} \Delta T.
\label{jleq}
\ee
In the above, $\Delta V = V_c - V_h >0$, whereas $\Delta T = T_h -T_c >0$. 
Since $J = I/A$ and the resistance of the material is $R = \rho l/A$,
we can write Eq. (\ref{jleq}) as
\be
I = -\frac{\Delta V}{R} + \frac{\alpha}{R} \Delta T.
\label{curvt}
\ee
which can be rewritten in the form 
\be
J_1 = \frac{T_c}{R} X_1 + \frac{\alpha T_c T_h}{R} X_2,
\label{linj1}
\ee
following the definitions in Eqs. (\ref{j1mac}) and (\ref{j2mac}).

Thus, we note that the flux $J_1 \equiv I$ is {\it linear} in the discrete forces 
$X_1$ and $X_2$.

Similarly, from Eqs. (\ref{ons2})
and (\ref{lij}), the local thermal flux inside the thermoelectric material is 
given by:
\be
J_{Q}^{} (x) = -\frac{\alpha T(x)}{\rho} \frac{\partial V}{\partial x} 
- \left(\frac{\alpha^2 T(x)}{\rho} + \kappa \right) \frac{\partial T}{\partial x},
\label{jetq}
\ee
which can be cast in the following form \cite{Apertet2013B}:
\be
J_{Q}^{}(x) = \alpha T(x) J + \frac{\kappa \Delta T}{l} -\frac{\rho(l-2x)J^2}{2}.
\label{jqx}
\ee
Then, thermal currents at the end points of the thermoelectric material
are evaluated to be  
\bea
\dot{Q}_h^{} &=& \alpha T_{h} I +K \Delta T-\frac{1}{2}RI^{2}, 
\label{hflux}\\
\dot{Q}_c^{} &=& \alpha T_{c} I +K \Delta T + \frac{1}{2} RI^{2} \label{cflux},
\eea
where $\dot{Q}_c^{} = J_Q(l) A$.
Also,  $K= \kappa A/l$ is the thermal conductance. Then, the total power 
output of thermoelectric generator, $P = \dot{Q}_h^{} - \dot{Q}_c^{}$
is given by: $P = \alpha \Delta T I - R I^2$.

We close this subsection by expressing the thermal flux $J_2 \equiv \dot{Q}_h$ 
in terms of the forces ($X_1, X_2$). Substituting for $I= J_1$ from Eq. (\ref{linj1})
into Eq. (\ref{hflux}), we obtain
\begin{widetext}
\be
J_2 =\frac{\alpha T_c T_h}{R} X_1 +\left(\frac{\alpha^2 T_h}{R}+K \right) T_h T_c X_2
-\frac{T_c^2}{2R}X_1^2-\frac{\alpha^2 T_h^2 T_c^2}{2R}X_2^2
-\frac{\alpha T_hT_c^2}{R}X_1 X_2.
\label{linj2}
\ee
\end{widetext}
Thus, we have expressed the fluxes ($J_1, J_2$) 
in terms of the global affinities ($X_1, X_2$).
Whereas $J_1$ is linear in the global forces, $J_2$ is explicitly non-linear
in these forces. Alternately, nonlinearity lies in the quadratic 
dissipation term in Eq. (\ref{hflux}), 
due to Joule heating \cite{Apertet2013B}. 
\subsection{Global kinetic coefficients}
We may formally define a set of kinetic coefficients
by assuming expansion of the fluxes in terms of 
the global affinities, as follows. We write
\bea
J_1  &=& {\cal L}_{11} X_1 + {\cal L}_{12} X_2,
\label{j1glo} \\
J_2 &=& {\cal L}_{21} X_1 + {\cal L}_{22} X_2 + {\cal O}[X_{j}^{2}].
\label{j2glo}
\eea
Then, comparing Eq. (\ref{j1glo}) with (\ref{linj1}), and 
Eq. (\ref{j2glo}) with (\ref{linj2}), we obtain
\bea
{\cal L}_{11} &=& \frac{T_c}{R}, \nonumber \\ 
{\cal L}_{12} &=& \frac{\alpha T_c T_h}{R} = {\cal L}_{21}, \label{callij} \\
{\cal L}_{22} &=& T_c T_h \left( \! \frac{\alpha^2 T_h}{R} + K \! \right). \nonumber 
\eea
Additionally, we have higher-order coefficients related to the quadratic terms in Eq. (\ref{linj2}).
The above kinetic coefficients may be compared with their local
counterparts in Eq. (\ref{lij}). Now, here also we observe the 
 equality of the cross-coefficients. Recall, that the corresponding
equality at the local level can be argued on the basis of microscopic
time-reversibility. Interestingly, the equality at the macroscopic
level can also be traced to the equality at the local level, as we
show below. 

From Eq. (\ref{ons2}), we have
\be
\vec{J}_Q(x) = \frac{L_{21}}{e L_{11}}\vec{J} - \frac{D}{L_{11}T^2} \vec{\nabla} T,
\label{jqx2}
\ee
where $D = L_{11}L_{22} - L_{12}L_{21}$. If we do not invoke Eqs. (\ref{lij}), 
then, equivalent to Eq. (\ref{hflux}), we can write 
\be
\dot{Q}_h = \frac{L^{'}_{21}}{e L^{'}_{11}} I + \frac{D'A}{L^{'}_{11}T^{2}_{h}l} \Delta T
- \frac{T_h\; l}{2e^2 L^{'}_{11}A} I^2,
\label{qhlij}
\ee
where the primed Onsager coefficients are evaluated 
at $x=0$, or $T = T_h$. Alternately, in terms of the forces $X_i$, we can write
$J_2 \equiv \dot{Q}_h$ as
\be
J_2 = \frac{L^{'}_{21} T_c}{e L^{'}_{11}R}X_1 + T_c T_h \left(
\frac{D'A }{L^{'}_{11}T^{2}_{h}l} + \frac{\alpha L^{'}_{21} T_c}{e L^{'}_{11}R}
\right) X_2 + 
{\cal O}[X_{j}^{2}].
\label{j2lp}
\ee
Comparing the above equation with Eq. (\ref{j2glo}), we identify
${\cal L}_{21} = {L^{'}_{21} T_c}/{e L^{'}_{11}R}$,
which turns out to be equal to ${\cal L}_{12}$, upon substituting
from Eq. (\ref{lij}). Thus, we see that the equality ${\cal L}_{21} 
= {\cal L}_{12}$ is consistent with the equality of Onsager coefficients
($L_{21} = L_{12}$) at the local level.
\section{Comparison with MNLIT model}
As mentioned in the Introduction, 
the so-called MNLIT model \cite{Izumida2012} assumes the following 
extended Onsager relations at the {\it global} level:
\bea
\bar{J}_1  &=& \bar{\cal L}_{11} X_1 + \bar{\cal L}_{12} X_2,
\label{j1gloz} \\
\bar{J}_2  &=& \bar{\cal L}_{21} X_1 + \bar{\cal L}_{22} X_2 
- \gamma_h \bar{J}_{1}^{2},
\label{j2gloz}
\eea
where $\gamma_h >0$ specifies the strength of dissipation into the hot reservoir.
The Onsager-like kinetic coefficients $\bar{\cal L}_{ij}$ are required to satisfy
specific conditions in order to satisfy the second law at the 
global scale, $\dot{S} = \bar{J}_1 X_1 + \bar{J}_2 X_2 >0$.

Now, by inspection, one can notice that the thermoelectric generator
at the global level of description, seems like a special case of
the MNLIT model, with the kinetic coefficients
$\bar{\cal L}_{ij} = {\cal L}_{ij}$ as given by Eq. (\ref{callij}), and 
$\gamma_h = R/2$. However, important distinctions between the two frameworks
need to be emphasized. First,
in the thermoelectric case, the global flux-force relations
are not postulated {\it per se}, but they emerge naturally when we scale
up the description from a local to the global level, as shown in the
previous sections. On the other hand,  the flux-force relations
in the form of Eqs. (\ref{j1gloz}) and (\ref{j2gloz}) are the premise
of MNLIT model. 
Secondly, the reciprocity of cross-coefficients,
$\bar{\cal L}_{21} = \bar{\cal L}_{12}$, is also something which is assumed
in the MNLIT model. We have seen in the above that this feature
is automatically satisfied for the thermoelectric model, and it is 
derived from the Onsager reciprocity at the local level.
Thus, as has been emphasized earlier \cite{Apertet2013B, Izumida2012}, 
the MNLIT model is not derived from some microscopic model, whereas
the global description of thermoelectricity is
derived from the underlying local linear-irreversible model. 
Still, the MNLIT model may be useful for a 
macroscopic modelling of irreversible thermal machines
in the nonlinear regime \cite{Izumida2015}. 
\section{Concluding remarks}
In the above, we have analyzed the flux-force relations for 
a thermoelectric generator at global level, by scaling up from the local,  
linear-irreversible thermodynamic framework of Onsager and Callen.
An important step is the identification of global affinities or
thermodynamic forces and fluxes from the bilinear expression
for the rate of total entropy generation. The flux-force
relation is linear in case of the driven flux (electric current)
while it is quadratic in the global forces for the driver flux 
(hot thermal flux). The nonlinear dissipation terms are known 
to be due to Joule heating. Further, these flux-force relations also yield
expressions for the effective kinetic coefficients. We
observe the equality of cross-coefficients and show that
this is derived from the Onsager reciprocity at the local level.
Although, the traditional picture of thermoelectricity is 
adequate to explain the global features of power generation and so
on, we believe the global flux-force relations expressed 
in terms of discrete or global equivalent of the local 
forces, is also a valid depiction of the same phenomenon. 
Our analysis also clarifies the comparison with the 
other nonlinear phenomenological models such as MNLIT model.
One can similarly formulate the global relations in the 
presence of magnetic field which breaks the time-reversal
symmetry \cite{Apertet2013B}. As may be expected, 
the global cross-coefficents are also not equal in this case.
Finally, in the present paper, we have assumed ideal thermal contacts
between the thermoelectric material and the reservoirs.
In principle, one can also include the effect of finite-conductance
heat exchangers in the performance analysis \cite{Gordon1991}.
It will be interesting, albeit more involved, to investigate the 
global force-flux picture in the presence of internal as well 
as external irreversibilities.

\begin{acknowledgments}
JK is grateful to Indian Insitute of Science Education and Research 
Mohali for financial support in the form of Senior Research Fellowship. 
\end{acknowledgments}

\appendix
\section{Choice of fluxes and affinities}
We recall that the choice of fluxes and conjugate affinities is not unique
within the Onsager framework \cite{Callen1948}. Thus, 
if the particle current density $\vec{J}_N$ and thermal current density
$\vec{J}_Q$ are the fluxes of our description, then $ -(\vec{\nabla} \mu)/T$ 
and $\vec{\nabla}(1/T)$, respectively, are the affinities. On the other hand, 
if $\vec{J}_N$ along with the energy current density, defined 
as: $\vec{J}_E (x) = \vec{J}_Q(x) + \mu(x) \vec{J}_N$, are chosen as the fluxes, then the 
corresponding affinities are, $-\vec{\nabla} (\mu/T)$ and $\vec{\nabla}(1/T)$, respectively. 
Analogously, the rate of entropy production at local level is given by:
\be 
\dot{s} = -\vec{J}_N. \vec{\nabla} \left( \frac{\mu}{T} \right)  + \vec{J}_{E}^{}(x).
\vec{\nabla} \left(  \frac{1}{T} \right),
\label{app1}
\ee
The corresponding global rate of entropy production is then given by:
\be
\dot{\cal S} = J_N \left(\frac{\mu_h}{T_h}  - \frac{\mu_c}{T_c}  \right)
  + J_{E}^{}(0) \left(  \frac{1}{T_c} - \frac{1}{T_h}  \right),
  \label{app2}
\ee
which is in a bilinear form, expressed in terms of the corresponding fluxes 
and discrete forces.
\section{Thermoelectric refrigerator}
In this section, we study the thermoelectric device working as a refrigerator.
 In this case, thermal and work flows invert their directions relative to the case of 
 thermoelectric generator. 
 At the local level, the fluxes are the electric flux density $\vec{J}$ and 
 thermal flux density $\vec{J}_Q$, with conjugate affinities as $\vec{\nabla} V /T$ 
 and $-\vec{\nabla} (1/T) = \vec{\nabla} T/T^2$.
 Local power input p.u.a. is given by $P= - \vec{\nabla}. \vec{J_Q} = \vec{J}.\vec{\nabla} V$.
Now, the rate of local entropy production per unit volume, $\dot{s}=-\vec{\nabla}.(\vec{J}_Q/T)$
can be written as follows.
 \be 
 \dot{s} = \vec{J}. \left(  \frac{\vec{\nabla} V}{T}  \right) - \vec{J}_{Q}^{}(x).
 \vec{\nabla} \left(  \frac{1}{T} \right),
 \label{sdotR}
\ee
In this case, local force-flux relations are expressed as:
\be           
\left(
\begin{array}{c}
	\vec{J} \\
	\vec{J}_Q
\end{array}
\right)
=
\left(
\begin{array}{cc}
	e^2 L_{11} &   e L_{12} \\ 
	e L_{21}  &   L_{22} 
\end{array}
\right)
\left(
\begin{array}{c}
	{\vec{\nabla} V}/{T} \\
	\vec{\nabla} T/{T^2}
\end{array}
\right),
\label{ons3}
\ee
where the Onsager coefficients in terms of the properties of thermoelectric material 
have the same expressions as in Eq. (\ref{lij}).\\
Now, the rate of entropy production p.u.a along the 1-d thermoelectric material
is given as
\be
\dot{\cal S} = -\frac{J_{Q}^{}(l)}{T_c} + \frac{J_{Q}^{}(0)}{T_h},
\label{sdot2R}
\ee
which can be rewritten as
\be
\dot{\cal S} = J \left( \frac{\Delta V}{T_h}  \right)
+ J_{Q}^{}(l) \left(  \frac{1}{T_h} - \frac{1}{T_c}  \right).
\label{sdotmacR}
\ee
The rate of total entropy production, $\dot{S} = 
\dot{\cal S} A  \equiv {J}_1 X_1 + {J}_2 X_2$,
where 
\bea
{J}_1 &=& I, \qquad X_1 =  \frac{\Delta V}{T_h},
\label{j1macr}\\
{J}_2 &=& \dot{Q}_c, \quad X_2 = \frac{1}{T_h} - \frac{1}{T_c}.
\label{j2macr}
\eea  
The  above fluxes $J_i$ may be expressed in terms of macroscopic affinities 
$X_i$. By using Eqs. (\ref{lij}) and (\ref{ons3}), we get
\be
\vec{J} = \frac {\vec{\nabla} V}{\rho} + \alpha \frac{ \vec{\nabla} T}{\rho}
\ee 
Integrate it over the whole length, i.e, $x \in [0,l]$, we obtain
$I = {\Delta V}/{R} - \alpha {\Delta T}/{R}$, which can be written as
\be
J_1 = \frac{T_h}{R} X_1 + \frac{\alpha T_c T_h}{R} X_2.
\label{j1R}
\ee
Also, thermal flux $J_2$ in terms of the global forces is given as
\be
J_2 = \frac{\alpha T_h T_c}{R}X_1+\left(\frac{\alpha^2 T_c}{R}+K\right)T_h T_c X_2
+ {\cal O}[X_{j}^{2}]
\label{j2R}
\ee
It shows that $J_2$ is not a linear function of the global forces.
Next, we compare Eqs. (\ref{j1R}) and (\ref{j2R})  with  Eqs. (\ref{j1glo}) and 
(\ref{j2glo}) to get global kinetic coefficents:
\bea
{\cal L}_{11} &=& \frac{T_h}{R}, \nonumber \\ 
{\cal L}_{12} &=& \frac{\alpha T_c T_h}{R} = {\cal L}_{21}, \\
{\cal L}_{22} &=& T_c T_h \left(\frac{\alpha^2 T_c}{R} + K \right). \nonumber 
\label{callij2}
\eea
These coefficients of thermoelectric refrigerator  may be compared with 
Eq. (\ref{callij}) for thermoelectric generator. We observe the difference 
in the form of diagonal coefficients. Further, it also shows that cross-coefficients 
preserve their equality which emerges from time-reversal 
symmetry at the microscopic level.


\begin{thebibliography}{10}
%
\bibitem{Onsager1931}
L.~Onsager.
\newblock Reciprocal relations in irreversible processes. ii.
\newblock {\em Phys. Rev.}, 38:2265--2279, Dec 1931.

\bibitem{Callen1948}
H.~B. Callen.
\newblock The application of onsager's reciprocal relations to thermoelectric,
  thermomagnetic, and galvanomagnetic effects.
\newblock {\em Phys. Rev.}, 73:1349--1358, Jun 1948.

\bibitem{Domenicali1954}
C.~A. Domenicali.
\newblock Irreversible thermodynamics of thermoelectricity.
\newblock {\em Rev. Mod. Phys.}, 26:237--275, Apr 1954.

\bibitem{Callenbook1985}
H.~B Callen.
\newblock {\em Thermodynamics and an introduction to thermostatistics}.
\newblock New York : Wiley, 2nd ed edition, 1985.
\newblock Rev. ed. of: Thermodynamics. 1960.

\bibitem{DavidJou2008}
G.~Lebon and D.~Jou.
\newblock {\em Understanding non-equilibrium thermodynamics :}.
\newblock Springer,, Berlin, 2008.

\bibitem{Gordon1991}
J.~M. {Gordon}.
\newblock {Generalized power versus efficiency characteristics of heat engines:
  The thermoelectric generator as an instructive illustration}.
\newblock {\em American Journal of Physics}, 59:551--555, June 1991.

\bibitem{DiSalvo1999}
F.~J. DiSalvo.
\newblock Thermoelectric cooling and power generation.
\newblock {\em Science}, 285(5428):703--706, 1999.

\bibitem{RIFFAT2003}
S.~B. Riffat and X.~Ma.
\newblock Thermoelectrics: a review of present and potential applications.
\newblock {\em Applied Thermal Engineering}, 23(8):913 -- 935, 2003.

\bibitem{Shakouri2011}
A.~Shakouri.
\newblock Recent developments in semiconductor thermoelectric physics and
  materials.
\newblock {\em Annual Review of Materials Research}, 41:399--431, 2011.

\bibitem{Benenti2011}
G.~Benenti, K.~Saito, and G.~Casati.
\newblock Thermodynamic bounds on efficiency for systems with broken
  time-reversal symmetry.
\newblock {\em Physical Review Letters}, 106(23), 2011.

\bibitem{snyder2011}
G.~J. Snyder and E.~S. Toberer.
\newblock {\em Complex thermoelectric materials}.
\newblock World Scientific, 2011.

\bibitem{Goupil2011}
C.~Goupil, W.~Seifert, K.~Zabrocki, E.~Müller, and G.~J. Snyder.
\newblock Thermodynamics of thermoelectric phenomena and applications.
\newblock {\em Entropy}, 13(8):1481--1517, 2011.

\bibitem{Apertet2012B}
Y.~{Apertet}, H.~{Ouerdane}, C.~{Goupil}, and P.~{Lecoeur}.
\newblock {Irreversibilities and efficiency at maximum power of heat engines:
  The illustrative case of a thermoelectric generator}.
\newblock {\em \pre}, 85(3):031116, March 2012.

\bibitem{Apertet2013A}
Y.~Apertet, H.~Ouerdane, A.~Michot, C.~Goupil, and Ph. Lecoeur.
\newblock On the efficiency at maximum cooling power.
\newblock {\em EPL (Europhysics Letters)}, 103(4):40001, 2013.

\bibitem{Ouerdane2013}
H.~Ouerdane, C.~Goupil, Y.~Apertet, A.~Michot, and A.l Abbout.
\newblock {\em A Linear Nonequilibrium Thermodynamics Approach to Optimization
  of Thermoelectric Devices}, pages 323--351.
\newblock Springer Berlin Heidelberg, Berlin, Heidelberg, 2013.

\bibitem{Wohlman2014}
O.~Entin-Wohlman, J.-H. Jiang, and Y.~Imry.
\newblock Efficiency and dissipation in a two-terminal thermoelectric junction,
  emphasizing small dissipation.
\newblock {\em Phys. Rev. E}, 89:012123, Jan 2014.

\bibitem{BENENTI2017}
G.~Benenti, G.~Casati, K.~Saito, and R.~S. Whitney.
\newblock Fundamental aspects of steady-state conversion of heat to work at the
  nanoscale.
\newblock {\em Physics Reports}, 694:1 -- 124, 2017.

\bibitem{Jasleen2019}
J.~Kaur and R.~S. Johal.
\newblock Thermoelectric generator at optimal power with external and internal
  irreversibilities.
\newblock {\em Journal of Applied Physics}, 126(12):125111, 2019.

\bibitem{Ding2019}
Y.~Ding, Y.~Qiu, K.~Cai, Q.~Yao, S.~Chen, L.~Chen, and J.~He.
\newblock High performance n-type ag2se film on nylon membrane for flexible
  thermoelectric power generator.
\newblock {\em Nature Communications}, 10(1):841, Feb 2019.

\bibitem{CWelton}
H.~B. Callen and T.~A. Welton.
\newblock Irreversibility and generalized noise.
\newblock {\em Phys. Rev.}, 83:34--40, Jul 1951.

\bibitem{Apertet2013B}
Y.~{Apertet}, H.~{Ouerdane}, C.~{Goupil}, and P.~{Lecoeur}.
\newblock {From local force-flux relationships to internal dissipations and
  their impact on heat engine performance: The illustrative case of a
  thermoelectric generator}.
\newblock {\em \pre}, 88(2):022137, August 2013.

\bibitem{Caplan1965}
O.~Kedem and S.~R. Caplan.
\newblock Degree of coupling and its relation to efficiency of energy
  conversion.
\newblock {\em Trans. Faraday Soc.}, 61:1897--1911, 1965.

\bibitem{VandenBroeck2005}
C.~Van~den Broeck.
\newblock Thermodynamic efficiency at maximum power.
\newblock {\em Phys. Rev. Lett.}, 95:190602, Nov 2005.

\bibitem{Iyyappan2020}
I.~Iyyappan and R.~S. Johal.
\newblock Efficiency of a two-stage heat engine at optimal power.
\newblock {\em {EPL} (Europhysics Letters)}, 128(5):50004, jan 2020.

\bibitem{Izumida2012}
Y.~Izumida and K.~Okuda.
\newblock Efficiency at maximum power of minimally nonlinear irreversible heat
  engines.
\newblock {\em {EPL} (Europhysics Letters)}, 97(1):10004, jan 2012.

\bibitem{Izumida2015}
Y.~Izumida, K.~Okuda, J.~M.~M. Roco, and A.~Calvo Hern\'andez.
\newblock Heat devices in nonlinear irreversible thermodynamics.
\newblock {\em Phys. Rev. E}, 91:052140, May 2015.

\bibitem{Esposito2010}
M.~Esposito, R.~Kawai, K.~Lindenberg, and C.~Van~den Broeck.
\newblock Efficiency at maximum power of low-dissipation {Carnot} engines.
\newblock {\em Phys. Rev. Lett.}, 105:150603, Oct 2010.

\bibitem{Izumida2013}
Y.~Izumida, K.~Okuda, A.~Calvo Hern{\'{a}}ndez, and J.~M.~M. Roco.
\newblock Coefficient of performance under optimized figure of merit in
  minimally nonlinear irreversible refrigerator.
\newblock {\em {EPL} (Europhysics Letters)}, 101(1):10005, jan 2013.

\bibitem{Onsager1931b}
Lars Onsager.
\newblock Reciprocal relations in irreversible processes. ii.
\newblock {\em Phys. Rev.}, 38:2265--2279, Dec 1931.

\bibitem{Ioffe1957}
A.~F. Ioffe.
\newblock {\em Semiconductor thermoelements, and Thermoelectric cooling}.
\newblock Infosearch, ltd., 1957.

\end{thebibliography}

\end{document}